\documentclass[12pt]{iopart}
\usepackage{graphicx}
\usepackage{epsfig}
\usepackage{amssymb}
\usepackage[english]{babel}

\begin{document}

\title[Phase dynamics and IV-characteristics of two parallel stacks of coupled Josephson junctions]
{Phase dynamics and IV-characteristics of two parallel stacks of coupled Josephson junctions}
\author{Yu. M. Shukrinov$^{a,b}$, I. R. Rahmonov$^{a,c}$, A. Plecenik$^{d}$,  M. Grajcar$^{d}$, P. Seidel$^{e}$,   E. Il'ichev$^{f,g}$, W. Nawrocki$^{h}$}

\address{$^a$ BLTP, Joint Institute for Nuclear Research, Dubna, Moscow Region, 141980,  Russia}
\address{$^b$ Department of Theoretical Physics, International University of Dubna, Dubna,  141980, Russia}
\address{$^c$ Umarov Physical Technical Institute, TAS, Dushanbe, 734063 Tajikistan}
\address{$^d$ Department of Experimental Physics, Comenius University, Bratislava, Slovakia}
\address{$^e$ Institut f$\ddot{u}$r Festk$\ddot{o}$rperphysik, Jena, D-07743 Jena, Germany}
\address{$^f$ Leibniz Institute of Photonic Technology, P.O. Box 100239, D-07702 Jena, Germany}
\address{$^g$ Novosibirsk State Technical University, 20 Karl Marx Avenue, 630092 Novosibirsk, Russia}
\address{$^h$ Poznan University of Technology, Poznan, Poland}
\ead{shukrinv@theor.jinr.ru}

\date{\today}

\begin{abstract}
Two parallel stacks of coupled Josephson junctions are investigated to clarify  the physics of transitions between the rotating and oscillating states and their effect on the IV-characteristics of the system. The detailed study of phase dynamics and bias dependence of the superconducting and diffusion currents allows one to explain all features of simulated IV-characteristics and demonstrate the correspondence in their behavior. The coupling between JJ in the stacks leads to the branching of IV-characteristics and a decrease in the hysteretic region. The crucial role of the diffusion current in the formation of the IV-characteristic of the parallel stacks of coupled Josephson junctions is demonstrated. We discuss the effect of symmetry in a number of  junctions in the stacks and show a decrease of the branching in the symmetrical stacks. The observed effects might be useful for development of superconducting electronic devices based on intrinsic Josephson junctions.
\end{abstract}
\submitto{\SUST}
\maketitle

\section{Introduction}
A strongly anisotropic high-$T_c$ superconductor (HTSC) forms a natural stack of  Josephson junctions (JJ) and shows the intrinsic Josephson effect \cite{kleiner92}. A model for describing
the physical properties of intrinsic JJs  in HTSCs, including nonlinear effects and various nonequlibrium phenomena, is offered by a system of coupled JJs ~\cite{krasnov2011,kurter2011}.
It should be emphasized that the system of coupled JJs is a promising object in the HTSC electronics, which has been extensively studied in recent years ~\cite{benseman11,koshelev2010,pfeiffer2008,yurgens00,grib2014}. In particular, broad possibilities for various applications are offered by the recently discovered coherent electromagnetic radiation of noticeable power in a terahertz frequency range, that is generated by the intrinsic JJs~\cite{ozyuzer07}. An interesting aspect of this phenomenon is that the emission is related to a certain region on the current-voltage characteristic (IV-characteristic), namely, that it corresponds to the parametric resonance (PR) in the JJ system ~\cite{sm-prl07,sms-prb08}.
Numerical modeling of the IV-characteristics predicts important properties of the system of coupled hysteretic intrinsic JJ (IJJ)  with McCumber parameters from 4 to 25 or even higher and  describes its phase dynamics in the branching (hysteretic) region of interest ~\cite{sr-jetpl10}.

For some applications it is interesting to consider the circuits which consist of two stacks of coupled JJ connected in parallel shown in Fig.~\ref{schema}(a). In the simplest case N1=N2=1, this is the well-known superconducting quantum interference device (DC-SQUID).  The main advantage of these devices based on HTSC is a possibility to operate at liquid nitrogen temperature (77 K) in comparison with the traditional SQUIDs operating at liquid helium temperature (4,2 K).  Modern HTSC DC SQUIDs are usually based on so-called grain-boundary junctions~\cite{ilichev96}. In this case, junctions are formed by the epitaxial grow of the HTSC films on the substrates with a certain topology and Josephson current flows parallel to the substrate surface (ab direction). This way the junction interface is not under control and the faceted nature (meandering) of the grain-boundary provides the spread of the junction parameters. Moreover, for the high angle of the grain misorientation, this type of junctions is described by the model of strongly inhomogeneous Josephson junctions \cite{ilichev99}.

For DC-SQUIDs the appearance of resonance features corresponding to the inductance L and capacitance C of the circuit like separate rc-branches was reported \cite{l1,l2,l3}. Additional strong influence of the external magnetic field leads to so-called ``beating branches''. Two junction SQUID can be extended to symmetric as well as asymmetric multi-junction loops, see \cite{darula95,x,y} and references therein. This type of circuits have a very complex dynamics. Multi-junction SQUIDs based on IJJs and  Shapiro steps in a DC-SQUID with multiple identical junctions in each arm were investigated.\cite{z1,z2,z3}

Different electronic devices based on IJJ, including SQUIDs and voltage standard arrays, are possibly work at liquid nitrogen temperature \cite{yurgens00,wang03}. One advantage can be to replace single artificial JJs by IJJ as a series connection. In the case of voltage standard as well as for radiation sources in the THz region the performance can be enhanced by a larger number of IJJ distributed in different coupled stacks \cite{wang03,wang01}. Tunable oscillators at THz frequencies from IJJ are of high interest for applications \cite{welp13,delfanazari13,delfanazari14}. To get a coherent radiation output, the stacks have to be synchronized\cite{lin14}. Multi-junction SQUIDs are very interesting, too \cite{kim04,irie05,okano06,lin14}, e.g. because of higher voltage signals.

An interface of artificial Josephson junctions with current flowing perpendicular to the substrate surface ($c$-direction), similar to the conventional technology for low temperature superconductors, can be controlled better. However, due to the extremely short coherence length of HTSC materials, such type of junctions is difficult to realize. Even junctions between HTSC and low-temperature superconductor in the $c$-direction usually having a small critical current density and low $I_{c}R_{n}$ (here $I_{c}$ is the critical current and $R_{n}$ is the normal resistance of a junction) product, make them useless for applications \cite{komissinski02}. Therefore, to use an intrinsic ("natural") Josephson junction is an alternative option. In the frame of this approach, a nonlinear element is supposed to be a stack of junctions. However, a direct use of these stacks faces a problem of intensive branching of their current voltage characteristics leading to the complexity of analysis of the data and preventing their application for magnetometery. So, a great challenge is to develop the principles and methods to exclude such switching of Josephson junctions and remove the main barrier on the way of creation of reliable SQUIDs based on this type of stacks. The coupling between junctions leads to the creation  of the rotating (the time average of $\frac{\partial\varphi(l)}{\partial t}$ is constant and that of $\sin(\varphi(l))$ is zero, $\varphi(l)$ is the phase difference of $l$th JJ) and oscillating  (the time average of $\frac{\partial\varphi(l)}{\partial t}$ is zero and that of $\sin(\varphi(l))$ is constant) states~\cite{matsumoto99}. The oscillating state (O-state) is one of the new elements which appear in IJJ in comparison with single JJ. The O-state can be realized if the number of junctions in the stack is more than two. In the framework of the present paper we make the first step in this direction and for the first time present the detailed study of the current voltage characteristics branching mechanism related to the switching of Josephson junctions in the stacks between the states with rotating and oscillating phases.

There is a series of experimental papers demonstrating the creation of SQIDs based on the stacks of IJJ in layered superconductors. Kim et. al.~\cite{kim04} made a SQUID based on IJJ (IJJ-SQUID) of BSCCO wiskers. Also a group of A. Irie~\cite{irie05, okano06} created an IJJ-SQUID and demonstrated the modulation of voltage by variation of applied magnetic flux. These results prove the possibility to create IJJ-SQUIDs. The theoretical background of IJJ-SQUID was made by V. M. Krasnov \cite{krasnov02}. However they could not realize the collective behavior of IJJ in the stacks. The main reason was the intensive switching of Josephson junctions in the stacks between the states with rotating and oscillating phases. These transitions lead to the branching of current voltage characteristics and make a  barrier on the way of creation of IJJ-SQUIDs.

On the other hand, an interesting application of a simple model describing the dynamics of the intrinsic Josephson junction stacks  in the presence of a hot spot by two parallel arrays of pointlike Josephson junctions and an additional shunt resistor in parallel was presented in Ref.\cite{gross13} The used model allowed one to explain the experimental results on the linewidth of terahertz radiation emitted from intrinsic Josephson junction stacks.

In this paper, we have presented the results of simulation of two parallel stacks with a different number of JJs. To clarify the physics of switching between different states, we study the system with a small number of JJs in the stacks. In comparison to the well--known capacitively coupled Josephson junction (CCJJ) model \cite{koyama96} we use the model with a diffusion current (CCJJ+DC model)  \cite{sms-physC06}. We have demonstrated that taking into account the diffusion current makes it possible to escape the  branching of IV-characteristics with an equal number of JJs in the stacks. We clearly show the switching mechanism by analysis of superconducting current and IV-characteristics of all JJs in both stacks. The comparison of the results for CCJJ and CCJJ+DC models stresses the importance of diffusion current in the switching processes.

The paper is organized as follows. In Sec. II, we introduce the model and describe the simulation procedure and parameters of simulations. A general consideration of the IV-characteristics for stacks with a different number of Josephson junctions is presented in Sec. III. In Sec. IV we discuss the case with one and three junctions in the stacks and show all transitions happen with the junctions with increasing and decreasing bias current.  The case with two and three junctions is analyzed in Sec. V. In Sec. VI, we present the symmetrical case with three junctions in both stacks. We compare and explain the results for CCJJ and CCJJ+DC models.  Finally, Sec. VII concludes the paper.
\section{Model and method}
\begin{figure}[htb]
 \centering
\includegraphics[height=45mm]{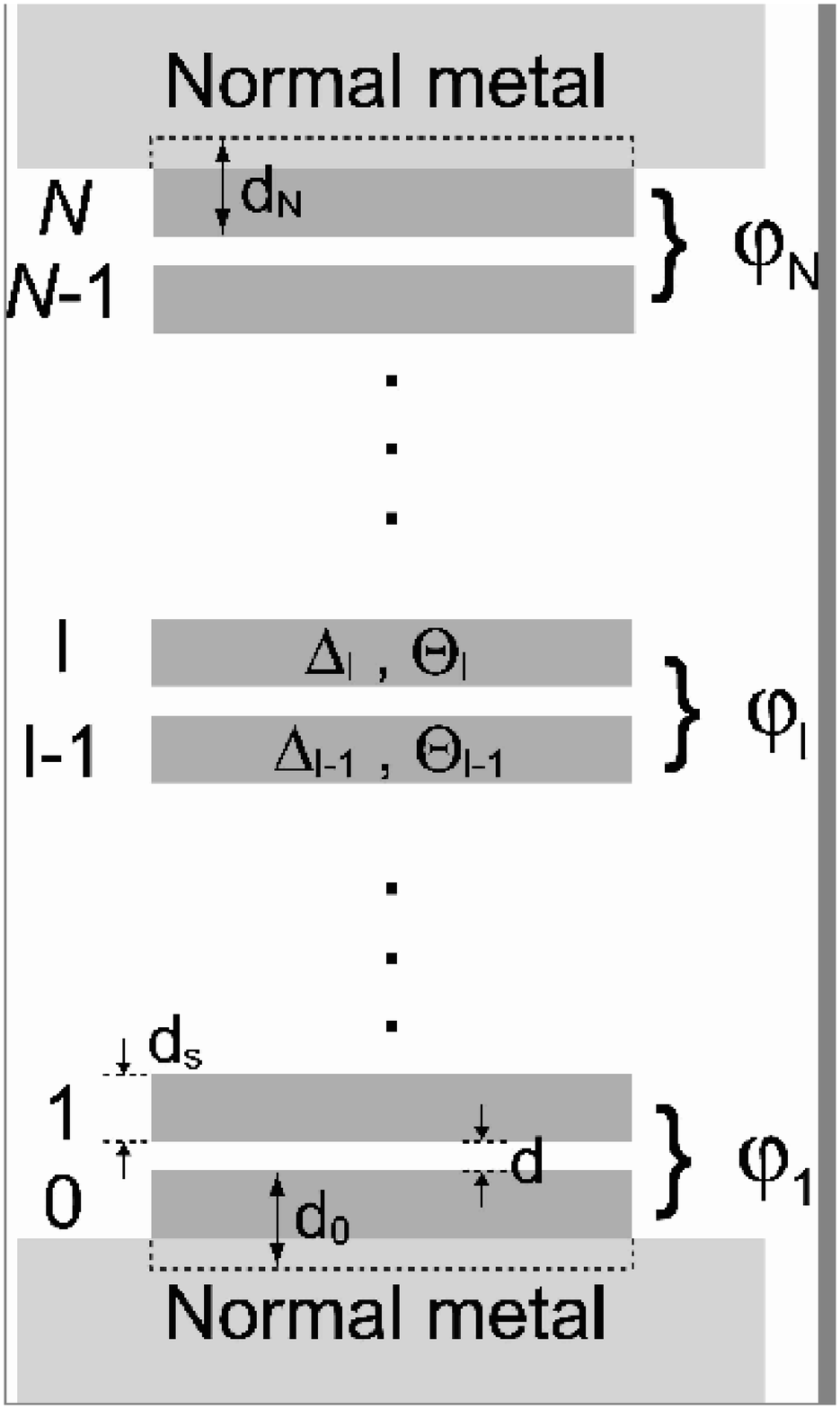}\hspace{0.5cm}\includegraphics[height=45mm]{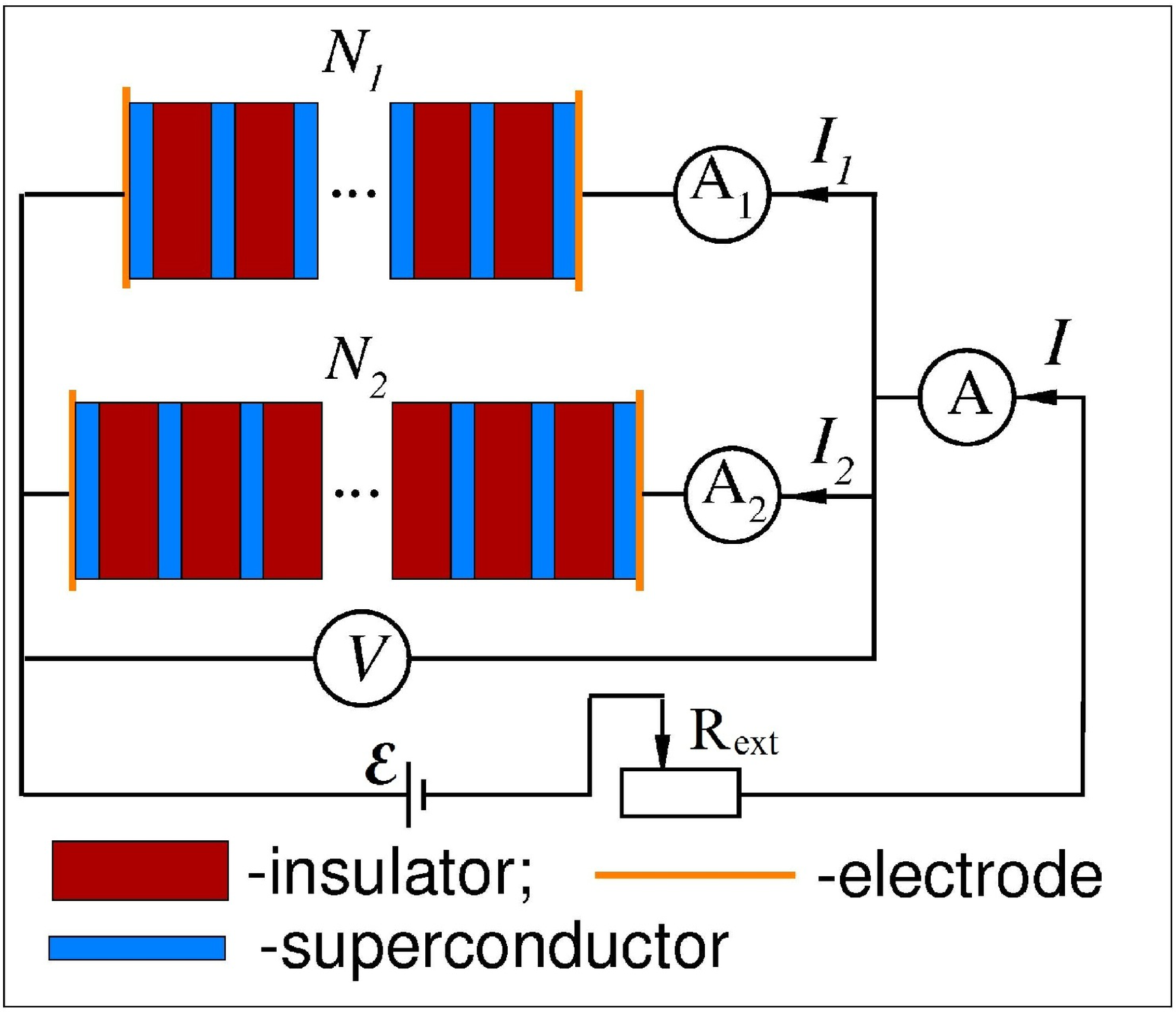}
\caption{(a) Schematic diagram of a system of intrinsic JJs in
HTSCs (see text); (b) Scheme of parallel connections of two stacks with $N_{1}$ and $N_{2}$ JJs}  \label{schema}
\end{figure}

A system of $N$ + 1 superconducting layers in an anisotropic HTSC, which is characterized by the order parameter $\Delta_l(t) = \vert \Delta \vert \exp(i\theta_l(t))$ with the time dependent phase  $\theta_l(t)$, comprises $N$ Josephson junctions \cite{kleiner92}. Figure~\ref{schema}(a) shows a schematic diagram of this layered system.

Superconducting layers, which are numbered by $l$ running from 0 to $N$ and characterized by the time-dependent order parameters with moduli $\Delta_{l}$ and phases $\theta_{l}$, form a system of IJJs with phase differences $\varphi_{l}=\theta_{l}-\theta_{l-1}$;  $d_{s}$ and $d$ are the thicknesses of the superconducting and dielectric layers, respectively.

The thickness of superconducting layers (about 3 {\AA}) in an HTSC is comparable with the Debye length $r_D$ of electric charge screening. Therefore, there is no complete screening of the charge in the separate layers, and the electric field induced in each IJJ penetrates into the adjacent junctions. Thus, the electric neutrality of superconducting layers is dynamically broken and, in the case of the alternating current (ac) Josephson effect, a capacitive coupling appears between the adjacent junctions  \cite{koyama96}. The absence of complete screening of the charge in the superconducting layer leads to the formation of a generalized scalar potential  $\Phi_l$ of the layer, which is defined in terms of the scalar potential  $\phi_l$ and the derivative of phase $\theta_l$ of the superconducting order parameter as follows: $\Phi_l(t)=\phi_l-V_0\frac{d\theta_{l}}{dt}$, where $V_0=\hbar\omega_p/(2e)$, $\omega_{p}=\sqrt{2eI_c/\hbar C_j}$ is the plasma frequency, $I_c$ is the critical current, and $C_j$ is the capacitance of JJ. The generalized scalar potential is
related to the charge density  $Q_l$  on the superconducting layer  $Q_l=-\frac{1}{4\pi r^{2}_D}\Phi_l$ \cite{koyama96, ryndyk98}.
The existence of a relationship between the electric
charge  $Q_l$  of the lth layer and the generalized scalar
potential  $\Phi_l$ of this layer reflects the nonequilibrium
nature of the ac Josephson effect in layered HTSCs  \cite{ryndyk98}.

When an external current passes through the stack of coupled Josephson junctions, the system appears under nonequilibrium conditions \cite{ryndyk98}. The Josephson relation is generalized in this case. Also the diffusion contribution to the quasiparticle current arises due to the generalized scalar potential difference
\begin{eqnarray}
I_{dif}^l=\frac{\Phi_{l} - \Phi_{l-1}}{I_c R_j} = - \frac{(Q_{l} - Q_{l-1})}{4\pi r^{2}_DI_c R_j }=
-\frac{(Q_{l}-
Q_{l-1})}{2e^2N(0)I_c R_j }= \nonumber \\
=\beta \dot{\varphi_l}-\beta V_l=-\alpha\beta(V_{l+1}+V_{l-1}-2V_{l}) \hspace{2cm} \label{dif-current}
\end{eqnarray}
where $R_j$ is the resistance of the Josephson junction in the resistive state and  $N(0)$ denotes the density of states on the Fermi level. The structure of the IV-characteristics in the capacitively coupled Josephson junction model with diffusion current (CCJJ+DC model) is equidistant \cite{sms-physC06} in agreement with the experiments~\cite{iso-apl08}. In Ref.~\cite{sr-jetp12}, it was demonstrated that the diffusion current plays an important role in the formation of the IV-characteristics, particularly, determining the width of the hysteretic region.

To investigate the phase dynamics of IJJ, we use the one-dimensional CCJJ+DC model with the gauge-invariant phase differences $\varphi_l(t)$  between $S$-layers $l$ and $l+1$ in the presence of electromagnetic irradiation described by the system of equations:
\begin{equation}
\label{syseq} \left\{\begin{array}{ll} \displaystyle\frac{\partial \varphi_{l}}{\partial
t}=V_{l}-\alpha(V_{l+1}+V_{l-1}-2V_{l})
\vspace{0.2 cm}\\
\displaystyle \frac{\partial V_{l}}{\partial t}=I-\sin \varphi_{l}-\beta\frac{\partial \varphi_{l}}{\partial t} + A\sin\omega_R t + I_{noise}
\end{array}\right.
\end{equation}
where $t$ is dimensionless time normalized to the inverse plasma frequency $\omega^{-1}_p$, 
$\displaystyle\beta=\frac{1}{R_{j}}\sqrt{\frac{\hbar}{2eI_{c}C_{j}}}=\frac{1}{\sqrt{\beta_{c}}}$,  $\beta_{c}$ is the McCumber parameter, $\alpha$  gives the coupling between junctions \cite{koyama96},
$A$ is the amplitude of the radiation.

For simplicity, we assume that the parameters of all junctions are equal. In a real case, there will be some spread in junction parameters which gives limited locking ranges for synchronization, see e.g.~\cite{seidel01}, but we like to demonstrate the main effects on the IV-characteristics here.

To find the IV-characteristic of the stack of intrinsic JJs, we solve the system of nonlinear second-order differential equations (1) using the fourth order Runge-Kutta method. In our simulations, we measure the voltage in units of $V_0$, the frequency in units of $\omega_{p}$, the bias current $I$ and the amplitude of radiation $A$ in units of $I_c$. We note that different kinds of couplings between junctions, like inductive
coupling in the presence of magnetic field \cite{bul96}, capacitive \cite{koyama96, machida99}, charge-imbalance \cite{ryndyk} and phonon \cite{helm, shu-sar02} couplings determine a variety of IV-characteristics observed in HTSC. The influence of these couplings  on the parametric resonance in the system is still an open problem.

The important information concerning the resonance features of intrinsic JJ in HTSC can be obtained by detailed investigation of the charge dynamics of superconducting layers. To study the time dependence of the electric charge in the $S$-layers, we use the Maxwell equation div$ (\varepsilon\varepsilon_0 E) = Q$, where $\varepsilon$ and
$\varepsilon_0$ are relative dielectric and electric constants, respectively. The charge density $Q_l$ in the $S$-layer $l$ is proportional to the difference between the voltages $V_{l}$ and $V_{l+1}$ in the neighbor insulating layers $Q_l=Q_0 \alpha (V_{l+1}-V_{l})$,
where $Q_0 = \varepsilon \varepsilon _0 V_0/r_D^2$.  The details of the simulation procedure are presented in Refs. \cite{sms-prb08,smp-prb07}.

The electrical scheme of the two parallel connected stacks is presented in Fig.\ref{schema}(b). The first stack consists of $N_{1}$ IJJ, the second one includes $N_{2}$ IJJ. The total current through the system is the sum of currents through each stack  $I=I_{1}+I_{2}$.

To describe the dynamics of coupled JJ in the stacks, we use the CCJJ+DC model \cite{sm-prl07,sr-jetp12}. In the framework of this model the corresponding currents  $I_{1}$ and $I_{2}$  are determined by the expressions
\begin{eqnarray}
\label{I_1_2}
I_{1}=C_{1}\frac{\partial U_{m}}{\partial t}+I_{c1}\sin \psi_{m}+\frac{\hbar}{2eR_{1}}\frac{\partial \psi_{m}}{\partial t}; \nonumber \\
I_{2}=C_{2}\frac{\partial V_{l}}{\partial t}+I_{c2}\sin \varphi_{l}+\frac{\hbar}{2eR_{2}}\frac{\partial \varphi_{l}}{\partial t},
\end{eqnarray}
\noindent where $U_{m}$ and $\psi_{m}$ determine the voltage and phase difference of the $m$-th JJ in the first stack, while $V_{l}$ and $\varphi_{l}$ in the $l$-th JJ of the second stack. The values of $C_{i}$, $I_{ci}$ and $R_{i}$ with corresponding indices $i=1,2$ express the capacitance, critical current and resistance of JJ in the first and second stacks, respectively.
The generalized Josephson relation for the first and second stacks are written in the form
\begin{eqnarray}
\label{phase_t}
\displaystyle\frac{\hbar}{2e}\frac{\partial \psi_{n}}{\partial t}=U_{m}-\alpha_{1}(U_{m+1}+U_{m-1}-2U_{m}); \nonumber \\
\displaystyle\frac{\hbar}{2e}\frac{\partial \varphi_{l}}{\partial t}=V_{l}-\alpha_{2}(V_{l+1}+V_{l-1}-2V_{l}),
\end{eqnarray}
\noindent where $\alpha_{1}$ and $\alpha_{2}$ are determined by the coupling parameter between JJ in the first and second stacks, respectively.

Let us go to the dimensionless values,  normalizing time $t$ to the plasma frequency of JJ in the second stack $\omega_{p}=\sqrt{2eI_{c2}/(C_{2}\hbar)}$, voltage - to $V_{0}=\hbar \omega_{p}/(2e)$, external current $I$ - to the critical current  $I_{c2}$ of the JJ in the second stack. Rewriting equations (\ref{I_1_2}) and (\ref{phase_t}) and taking into account that, the sums of the voltages in the first and second stacks are equal to each other  $\sum^{N_{1}}_{n=1}U_{n}=\sum^{N_{2}}_{k=1}V_{k}$ we get the system of nonlinear equations :
\begin{eqnarray}
\label{system_eq1}
\left\{\begin{array}{ll}
\displaystyle\dot{\psi}_{m}=U_{m}-\alpha_{1}(U_{m+1}+U_{m-1}-2U_{m})
\vspace{0.1 cm}\\
\displaystyle\dot{\varphi}_{l}=V_{l}-\alpha_{2}(V_{l+1}+V_{l-1}-2V_{l})
\vspace{0.1 cm}\\
\displaystyle \dot{U}_{m}=(I-I_{c}\sin \psi_{m}-\mu\dot{\psi}_{m}
-\dot{V}_{l}-\sin \varphi_{l}-\beta\dot{\varphi}_{l})/C
\vspace{0.1 cm}\\
\displaystyle\dot{V}_{l}=\sum^{N_{2}}_{k=1} B^{-1}_{lk}
[N_{1}(I-\sin \varphi_{k}-\beta\dot{\varphi}_{k}) \\
-\sum^{N_{1}}_{n=1}(I_{c}\sin \psi_{n}+\mu\dot{\psi}_{n})]
\end{array}\right.
\end{eqnarray}
where $C=C_{1}/C_{2}$, $ R=R_{1}/R_{2}$ and $I_{c}=I_{c1}/I_{c2}$, $\beta=R_{2}^{-1}\sqrt{\hbar/(2eI_{c2}C_{2})}$ and $\mu=\beta/R$. The elements of the matrix $B$ are determined as $B_{lk}=C+N_{1}\delta_{lk}$, where $\delta_{lk}$--Kronecker' symbols. The system of equations (\ref{system_eq1}) are solved numerically by the Runge-Kutta fourth order procedure. As a result, we find the phase differences  $\psi_{m}(t)$, $\varphi_{l}(t)$ and votages $U_{m}(t)$, $V_{l}(t)$ as functions of time in the interval  $[0\div T_{max}]$.

To find averages of the voltages $\bar{U}_{m}$ and $\bar{V}_{l}$ at each step of bias current, we use
\begin{eqnarray}
\label{U-V}
\bar{U}_{m}=1/(T_{max}-T_{min})\int_{T_{min}}^{T_{max}}U_{m}(t)dt; \nonumber \\ \bar{V}_{l}=1/(T_{max}-T_{min})\int_{T_{min}}^{T_{max}}V_{l}(t)dt,
\end{eqnarray}
where $T_{min}$ and $T_{max}$ are the low and high limits of the averaging interval, respectively.
In the last expressions, the numerical integration has been done by the rectangular method. The total average voltages are calculated as
\begin{eqnarray}
\label{total-U-V}
\bar{U}=\sum^{N_{1}}_{m=1} \bar{U}_{m}; \nonumber \\
\bar{V}=\sum_{l=1}^{N_{2}}\bar{V}_{l},
\end{eqnarray}
To find IV-characteristics for separate stacks, i.e.  $\bar{U}(I_{1})$ and  $\bar{V}(I_{2})$, we calculate $I_{1}$ and $I_{2}$. As it is known, in the CCJJ+DC model through each JJ flow the superconducting current   $I_{s}$, quasiparticle  $I_{qp}$, diffusion $I_{dif}$ and displacement $I_{disp}$ currents. In the normalized units these currents for the first and second stacks are determined by:
\begin{eqnarray}
\label{currents}
I^{m}_{s1}=\sin\psi_{m}, \hspace{0.2cm} I^{m}_{qp1}=\mu U_{m}, \nonumber \\
I^{m}_{dif1}=\mu(\dot{\psi}_{m}-U_{m}),\hspace{0.2cm} I^{m}_{disp1}=\dot{U}_{m}; \nonumber \\
I^{l}_{s2}=\sin\varphi_{l},\hspace{0.2cm} I^{l}_{qp2}=\beta V_{l},\nonumber \\
I^{l}_{dif2}=\beta(\dot{\varphi}_{l}-V_{l}), \hspace{0.2cm} I^{l}_{disp2}=\dot{V}_{l}.
\end{eqnarray}
Using these expressions we find the currents $I_{1}(t)$ and $I_{2}(t)$ through each stack as a function of time. The calculated values of currents are averaged by the same scheme we used for averaging the voltage. As a result, we get the averaged values of the currents $\bar{I}_{1}$ and $\bar{I}_{2}$, which determine the IV-characteristics of each stack because the voltages in the both stacks are found already.

In our simulations we consider that all JJ in both the stacks are identical, i.e. $C=1$, $I_{c}=1$ and $R=1$. The calculations were made  at different values of the dissipation parameter $\beta=0.1, 0.2$, coupling parameter of the first stack $\alpha_{1}=0.1, 1$ and the second one $\alpha_{2}=0.1, 1$. We note that the qualitative results are not sensitive in this interval of values. The bias current changes with a step $\bigtriangleup I=0.0001$. The time domain consists of $[0\div 2000]$ units with a step $0.05$ and averaging usually done in the interval $[50\div2000]$. The direction of currents in the figures are shown by the arrows.

\section{Switching in the parallel stacks: general consideration}
The system presented in Fig.~\ref{schema}(a) looks very simple, but it actually demonstrates a very complex behavior and is characterized by many different parameters and dependences. Particularly, by  (i) total IV-characteristics giving voltage of the stacks as a function of the total bias current $I$, (ii) IV-characteristics of each stack giving voltage of the stack as a function of the corresponding current $I_1$ or $I_2$, (iii) IV-characteristics of each IJJ in the stacks. The system is also characterized by superconducting, quasiparticle, displacement and diffusion currents through the stacks and their dependence on the corresponding currents $I_1$, $I_2$ and $I$. All these characteristics are determined by the phase dynamics of the system, which in its  turn is determined by the time dependence of the phase differences and voltages in each JJ of the stacks.

\begin{figure}[htb]
 \centering
\includegraphics[width=80mm]{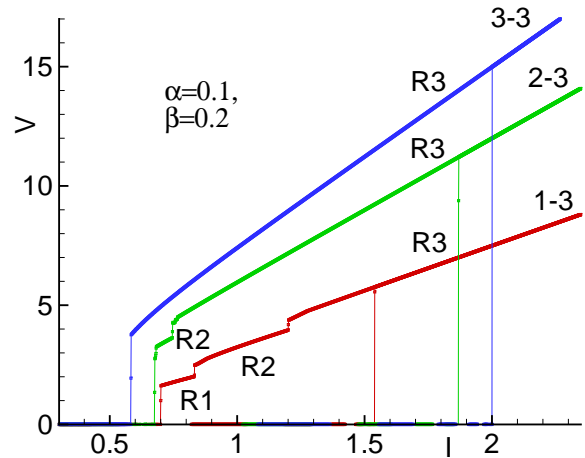}
\caption{One loop IV-characteristics for the different cases:  3-3 correspond to the case $N_{1}=3$, $N_{2}=3$; 2-3 - $N_{1}=2$, $N_{2}=3$; 1-3 - $N_{1}=1$, $N_{2}=3$. $R1$, $R2$, $R3$ indicate the branches with 1, 2 and 3 JJs in the rotating state, respectively.}   \label{cvc-all}
\end{figure}

First, we consider the one-loop IV-characteristics giving the voltage of the stacks as a function the of total bias current $I$. The results of simulations for the structures with a different $N_{1}=1, 2, 3$ and the same $N_{2}=3$ are presented in Fig.~\ref{cvc-all}. We see that The IV-characteristic in the case of 1-3-structure ($N_{1}=1$, $N_{2}=3$)  has three branches: ``R3'', ``R2'', and  ``R1''. The analysis of the averaged time derivative of the phase for each JJ of the stack shows that these branches reflect the states with a different number of rotating JJ in the second stack. The branch ``R3'' corresponds to the state with all three JJ in the R-state, the branch ``R2'' corresponds to the state with 2 JJ in the R-state and one JJ in the O-state, the branch ``R1'' corresponds to the state with 1 JJ in the R-state and 2 JJ in the O-state.

In the case of 2-3-structure the IV-characteristic has two branches: the first corresponds to the state with all three JJ in the R-state, the second one reflects the state with one JJ in the O-state and two JJ in the R-state. We will show below that these two junctions in the R-state have a transition to the superconducting state (zero voltage state) simultaneously.

In the case of the 3-3-structure the IV-characteristic has no inner branches at all: 3 JJ  do not demonstrate any intermediate  jumps to the $O$-states.  All JJ in the stacks undergo to the zero voltage state simultaneously at the same bias current.

Based on the presented results we might come to the conclusion that the system prefers to be in the state with both stacks having an equal number of JJ in the $R$-state with approximately the same value of voltage. In all considered cases the jumps to the zero voltage state happen from these states, escaping transitions of JJs to the $O$-states, i.e., the system prefers symmetry related to the voltage distribution between the stacks!  Let us make more detailed analysis of different cases.

\section{Case 1-3}
\begin{figure}[htb]
 \centering
\includegraphics[width=80mm]{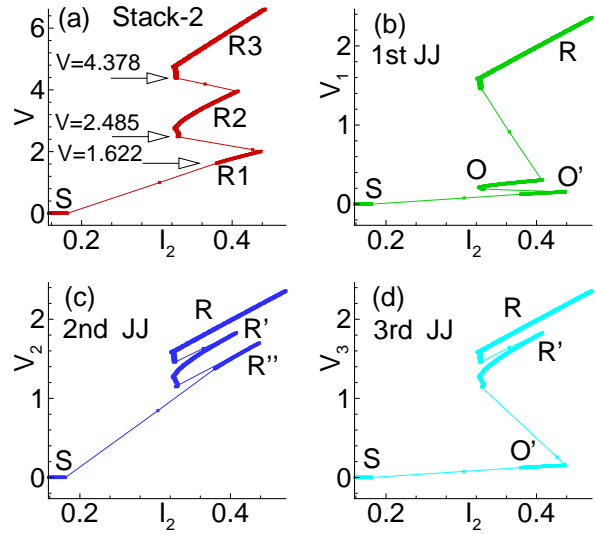}
\caption{(a) IV-characteristic of the second stack of 1-3-structure calculated with a decrease of the bias current; (b) IV-characteristics of the first JJ in the stack $V_1(I_2)$; (c) $V_2(I_2)$; (d) $V_3(I_2)$. The $O$ and $O'$ are the different  oscillating states and $R$, $R'$  and $R''$ are the different rotating states of JJs realizing in the stack.}  \label{n1-n3-stack2-down-JJ-cvc-v2}
\end{figure}

Consider the case when $N_1=1$ and $N_2=3$ (see Fig.~\ref{schema}(b)). First, we analyze the voltage in the second stack as a function of the corresponding current $I_2$, i.e. $V(I_2)$. Fig.~\ref{n1-n3-stack2-down-JJ-cvc-v2}(a) presents the IV-characteristics of the second stack calculated with a decrease of the current $I_2$.

We see here also three branches $R3$, $R2$, and  $R1$, related to the states of the stack with the corresponding number of junctions in the rotating state. Figures~\ref{n1-n3-stack2-down-JJ-cvc-v2}(b,c,d) show the IV-characteristics $V_j(I_2)$  of each junction in the stack. Follow the IV-characteristic by decreasing voltage we note that the first transition to the oscillating state happens with the first JJ, then third one, and lastly all three JJ undergo to the zero voltage state $S$. When the first JJ transits to the oscillating state $O$, the second and third ones transit to the rotating state $R'$. At the next transition, when the third JJ transits to another oscillating state $O'$, the first one also transits from $O$ to the $O'$-state, but the second JJ transits to another rotating state $R''$. At the third transition, all JJ undergo to the $S$ state. So we see clearly the correspondence between all transitions of JJs in the second stack.

\begin{figure}[htb]
 \centering
\includegraphics[width=80mm]{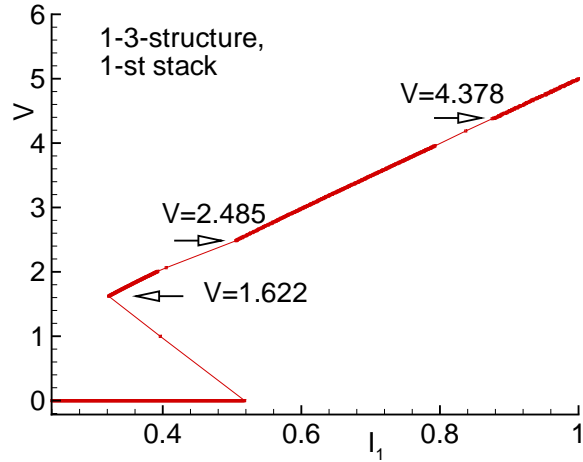}\hspace{0.5cm}
\caption{IV-characteristic of the first stack of the 1-3-structure  calculated with a decrease of the bias current. The arrows with numbers indicate the voltage values at the transitions.}  \label{n1-n3-stack1-down}
\end{figure}

These transitions in the second stack affect the behavior of JJ in the first one. Figure~\ref{n1-n3-stack1-down} presents the IV-characteristics of the first stack calculated also with a decrease of the bias current.
It reflects all jumps of JJs in the second stack by breaks in the curve. The corresponding jumps are shown by the arrows. At these points the frequency of JJ of the first stack changes abruptly. We see that the first jump happens at the voltage $V=4.378$ which  coincides with the corresponding value in Fig.~\ref{n1-n3-stack2-down-JJ-cvc-v2}(a). The same is correct concerning the other jumps. We see also that the first jump at $V=4.378$ happens at current $I_1$  much larger than $I_2$ at the corresponding jump presented in Fig.~\ref{n1-n3-stack2-down-JJ-cvc-v2}(a). The decrease in $I_1$ is compensated by the increase of $I_2$ observed in Fig.~\ref{n1-n3-stack2-down-JJ-cvc-v2}(a).
\begin{figure}[htb]
 \centering
\includegraphics[width=80mm]{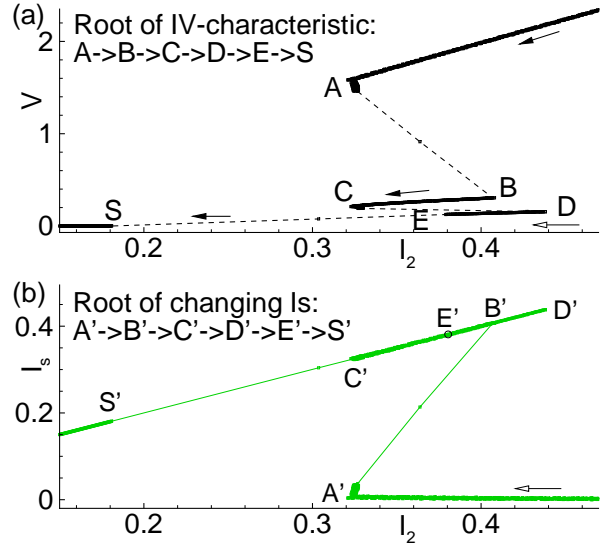}\hspace{0.5cm}
\caption{Comparison of the IV-characteristic (a) and superconducting current $I_s$ (b) of the first JJ in the second stack of the 1-3-structure calculated with a decrease of the bias current. The letters with prime refer to the $I_s$ curve.}  \label{cvc-Is-n1-n3-JJ1}
\end{figure}

To make clear the jumps increasing the current $I_2$ through the second stack and decreasing $I_1$ through the first one,  we study the corresponding changes of the superconducting current along the IV-characteristic. Figure~\ref{cvc-Is-n1-n3-JJ1} present the IV-characteristic (a) and  superconducting current $I_s$ (b) of the first JJ in the second stack of the 1-3-structure calculated with a decrease of the bias current. The letters mark the points of the IV-characteristic obtained by decreasing the total bias current $I$. The prime letters refer to the $I_s$ curve. The voltage decreases along the root $A \rightarrow B \rightarrow C \rightarrow D \rightarrow E \rightarrow S$ reflecting the transitions of JJs in the stack from the R-states  to the $O$-states. The jump at point $A$ correspond to the transition of the first junction. We see that at this jump  $A \rightarrow B$ on the IV-curve the decrease of $I_2$ is compensated by the increase of the superconducting current $ A' \rightarrow  B'$ on the $I_s$ dependence. Along the branch $BC$ the superconducting current decreases, following the root $B'C'$. Then a jump happens from C to D, i.e.,  to the state with two  oscillating JJ and we observe the corresponding increase of $I_s$ along $C'D'$. Then the superconducting current decreases following D'E'. From point $E$ transition to the superconducting state is realized.

\section{Case 2-3}
\begin{figure}[htb]
 \centering
\includegraphics[width=80mm]{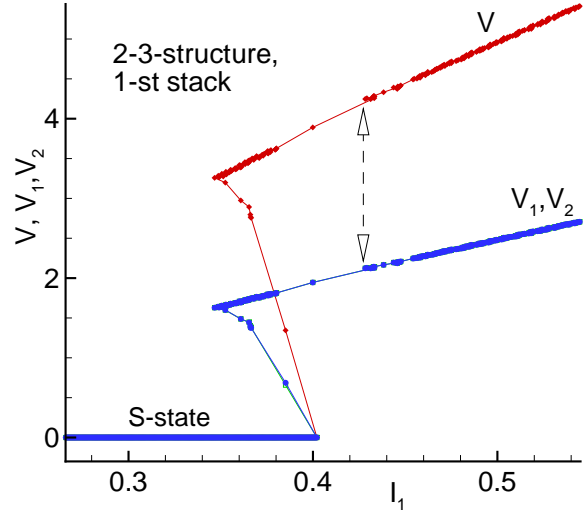}\hspace{0.5cm}
\caption{IV-characteristic of the first stack (V) and IV-characteristics of its JJs ($V_1$, $V_2$) of the 2-3-structure calculated with a decrease of the bias current. }  \label{n2-n3-stack1-down}
\end{figure}

Let us now consider the case $2-3$, i.e., two JJ in the first stack and three JJ in the second one. The corresponding IV-characteristic of the first stack (marked by letter V) and the IV-characteristics of its JJs ($V_1$, $V_2$) calculated with a decrease of the bias current are presented in Fig.~\ref{n2-n3-stack1-down}.
We see that the IV-characteristics of JJs in this stack coincide. Both stacks jump to the superconducting state simultaneously, stresses the conclusion in the end of Section III.

\begin{figure}[htb]
 \centering
\includegraphics[width=80mm]{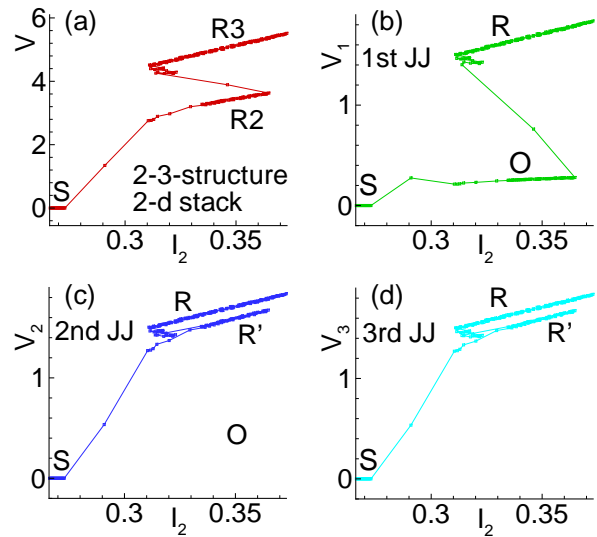}\hspace{0.5cm}
\caption{(a) IV-characteristic of the second stack of 2-3-structure calculated with a decrease of the bias current; IV-characteristics of each junction in the stack: (b) first JJ; (c) second JJ; (d) third JJ. }  \label{n2-n3-stack2-down-JJ-cvc-v2}
\end{figure}

Figure~\ref{n2-n3-stack2-down-JJ-cvc-v2}(a) presents the IV-characteristic of the second stack calculated also with a decrease of the bias current. We see two branches  $R3$ and  $R2$ related to the states of the stack with the corresponding number of junctions in the rotating state. Figures~\ref{n2-n3-stack2-down-JJ-cvc-v2}(b,c,d) show the IV-characteristics of each junction in the stack. It is clear from these figures that the first transition to the oscillating state $O$ happens with the first JJ. Both other JJs go to the rotating state $R'$. At the next transition all three junctions undergo to the superconducting state.

\section{Symmetrical case: the 3-3-structure}
From the previous consideration we have noted that there is an important reason for a different behavior of the 1-3- and  2-3-structures. The reason is related to a symmetry of the system.  The 2-3-structure has a possibility to jump to the symmetrical state with two JJs in both stacks. This state is stable and all four JJs undergo to the zero voltage state simultaneously with a decrease of the bias current.  This idea is also supported by simulation of the parallel connection of the stacks with 3 JJs in each of them. As it demonstrated in Fig.~\ref{cvc-all}, the corresponding IV-characteristics do not demonstrate any intermediate  jumps! The JJs in the stacks undergo to the superconducting state simultaneously at the same bias current.

\begin{figure}[htb]
 \centering
\includegraphics[width=80mm]{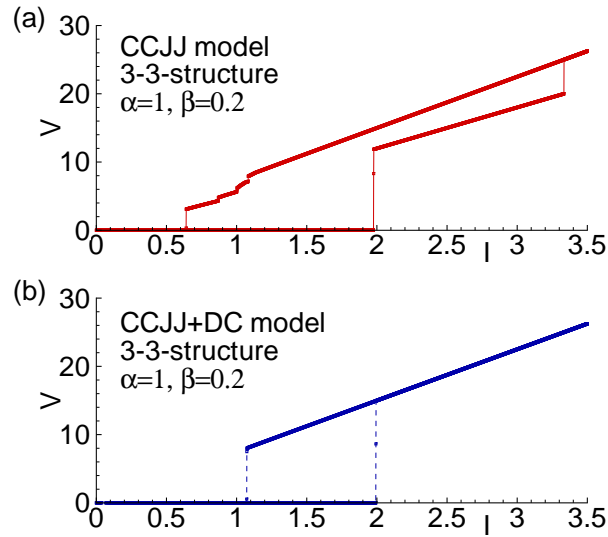}
\caption{(a) One loop IV-characteristics of the $3-3$-structure in CCJJ model; (b) The same for CCJJ+DC model. }  \label{compare-ccjj-ccjjdc}
\end{figure}

It was demonstrated in Ref.\cite{sr-jetp12} that the diffusion current  plays an important role in suppressing branching in the IV-characteristics of the single stack of coupled JJs. We consider that in our case of two parallel stacks the absence of branching in the IV-characteristics  has the same origin: It  is an effect of the diffusion current. To test this idea we simulate the IV-characteristics of parallel stacks in the framework of the CCJJ model that does not include a diffusion current and compare with the results of the CCJJ+DC model with the same parameters of the system. The results of simulations for the 3-3-structure  are presented in Fig.~\ref{compare-ccjj-ccjjdc}. To make features more pronounced we simulated the IV-characteristics at $\alpha=1$.

\begin{figure}[htb]
 \centering
\includegraphics[width=80mm]{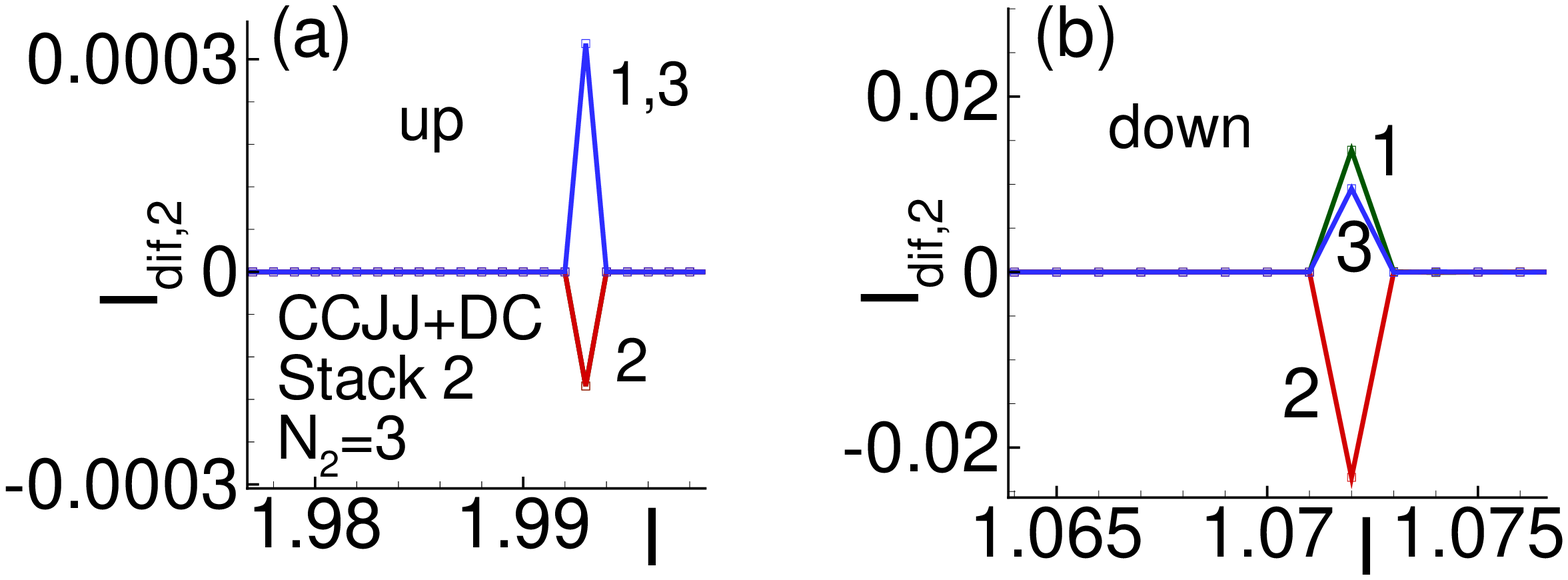}
\caption{(a) Diffusion current around $I_c$ of the $3-3$-structure in the CCJJ+DC model; (b) The same in the hysteretic region. }  \label{dif-currents}
\end{figure}

We see the branching of the IV-characteristic in the CCJJ model around the critical current and in the hysteretic region. However the  IV-characteristic of the CCJJ+DC model with the diffusion current  demonstrates a collective behavior of JJs, i.e., without transitions to any rotating and oscillating states.

The question then arises what suppresses such transition in this model.
The answer can be found in Fig.  9. The diffusion current changes its
direction  along the stack making the intermediate states unstable and forcing the system transit to the final state.\cite{sr-jetp12}  The behavior of the diffusion current around the critical value $I_c$ is shown in  Fig.~\ref{dif-currents}(a). We see that at the transition point the 1st and 3rd JJs have positive $I_{dif}$ but the 2nd one has negative value, i.e. opposite direction. The same features are observed in the hysteretic region presented in Fig.~\ref{dif-currents}(b). The diffusion current in the first stack reflects this behavior.

\section{Conclusions}

We studied the phase dynamics of two coupled parallel stacks of identical IJJ. We demonstrated that taking into account the diffusion current allows one to escape the branching of the IV-characteristics with an equal number of JJs in the stacks. We showed clearly the switching mechanism by analysis of the superconducting current and IV-characteristics of all JJs in both stacks. The comparison of the results for the CCJJ and CCJJ+DC models stress the importance of the diffusion current in the switching processes.

The coupling by dc as well as intrinsic rf currents plays an important role for synchronization of the junctions and locking ranges of common parts of the IV-characteristics. The coupled stacks of IJJ have potential for different applications like voltage standard, THz radiation sources and SQUID-like devices. Additional interest in IJJ arises because the new iron-based superconductors show intrinsic Josephson effects, too, see e.g. \cite{nakamura09,kashiwaya10,ayukawa14}.

\section{Acknowledgments}
The work was supported by the Heisenberg-Landau program, Bogoliubov-Infeld program and the JINR-Slovakia collaboration. We thank W. Kleinig, S. Dubnichka and W. Chmielovski for this support.


\section*{References}

\begin{thebibliography}{9}
\bibitem{kleiner92} Kleiner R, Steinmeyer F, Kunkel G and M$\ddot{u}$ller P 1992 \emph{Phys. Rev. Lett.} {\bf68} 2394
\bibitem{krasnov2011}Krasnov V M 2011 \emph{Phys. Rev. B} {\bf 83} 174517
\bibitem{kurter2011}Kurter C, Zhuravel A P, Ustinov A V and Anlage S M 2011 \emph{Phys. Rev. B} {\bf 84} 104515
\bibitem{benseman11}Benseman T M, Koshelev A E, Gray K E, Kwok W -K, Welp U, Kadowaki K, Tachiki M and Yamamoto T 2011 \emph{Phys. Rev. B} {\bf 84} 064523
\bibitem{koshelev2010} Koshelev A E 2010 \emph{Phys. Rev. B} {\bf 82} 174512
\bibitem{pfeiffer2008} Pfeiffer J, Abdumalikov Jr A A, Schuster M and Ustinov A V 2008 \emph{Phys. Rev. B} {\bf 77} 024511
\bibitem{yurgens00}Yurgens A A 2000 \emph{Supercond. Sci. Technol.} {\bf 13} R85
\bibitem{grib2014} Grib A, Seidel P 2014 \emph{Journal of Physics: Conference Series} {\bf 507} 042038
\bibitem{ozyuzer07}Ozyuzer L \emph{et al.} 2007 \emph{Science} {\bf 318} 1291
\bibitem{sm-prl07} Shukrinov Yu M and Mahfouzi F 2007 \emph{Phys. Rev. Lett.} {\bf 98} 157001
\bibitem{sms-prb08}Shukrinov Yu M, Mahfouzi F and Suzuki M 2008 \emph{Phys. Rev. B} {\bf 78} 134521
\bibitem{sr-jetpl10} Shukrinov Yu M and Rahmonov I R 2010 \emph{JETP Lett.} {\bf 92} 327; \emph{Pis'ma v ZhETF} {\bf 92} 364
\bibitem{ilichev96} Il'ichev E, Dorrer L, Hildebrandt G, Schmidl F, Zakosarenko V M, Seidel P  1996 \emph{Appl. Phys. Lett.} {\bf 68} 708
\bibitem{ilichev99}Il'ichev E, Zakosarenko V, IJsselsteijn R P J, Hoenig H E, Meyer H -G, Fistul M V, Mueller P 1999 \emph{Phys. Rev. B} {\bf 59} 11502
\bibitem{l1} Ben-Jacob E and Imry Y 1981 \emph{J. Appl. Phys.} {\bf 52} 6806
\bibitem{l2} Voss R F, Laibowitz R B, Broers A N, Raider S I, Knoedler C M and Viggiano J M 1981 \emph{IEEE Trans. Magnetics} {\bf 17} 395
\bibitem{l3} Schmidt W -D, Seidel P and Heinemann S 1985 \emph{Phys. Stat. Sol. (a)} {\bf 91} K155
\bibitem{x} Darula M, Seidel P, Busse F and Benacka S 1993 \emph{J. Appl. Phys.} {\bf 74} 2674
\bibitem{y} Grib A N, Seidel P and Darula M 1998 \emph{J. Low Temp. Phys.} {\bf 112} 323
\bibitem{darula95} Darula M, Beuven S, Siegel M, Darulova A and Seidel P 1995 \emph{Appl. Phys. Lett.} {\bf 67} 1618
\bibitem{z1} Krasnov V M 2002 \emph{Physica C} {\bf 368} 246
\bibitem{z2} Irie A and Oya G 2005 \emph{IEEE Trans Appl Supercond} {\bf 15} 813
\bibitem{z3} De Luca R and Romeo F 2005 \emph{J Appl Phys} {\bf 98} 073904
\bibitem{wang03} Wang HB, Chen J, Wu PH, Yamashita T, Vasyukov D, and M\"uller P 2003 \emph{Supercond. Sci. Technol.} {\bf 16} 1375
\bibitem{wang01} Wang HB, Wu PH, Yamashita T 2001 \emph{Appl. Phys. Lett.} {\bf 78} 4010
\bibitem{welp13} Welp U, Kadowaki K, Kleiner R 2013 \emph{Nature Photonics} {\bf 7} 702
\bibitem{delfanazari13}  Delfanazari K, Asai H, Tsujimoto M,  Kashiwagi T, Kitamura T, Yamamoto T, Sawamura M, Ishida K, Watanabe C, Sekimoto S, Minami H, Tachiki M,  Klemm R A, Hattori  T, Kadowaki K 2013 \emph{Optics Express} {\bf 21} 2171
\bibitem{delfanazari14} Delfanazari K, Asai H, Tsujimoto M,  Kashiwagi T, Kitamura T, Ishida K, Watanabe C,  Sekimoto S,  Yamamoto T, Minami H, Tachiki M,  Klemm R A, Hattori T, Kadowaki K 2014 \emph{J Infrared Milli Terahz Waves} {\bf 35} 131
\bibitem{lin14} Lin S-Z 2014 \emph{J. Appl. Phys.} {\bf 115} 173901
\bibitem{kim04}Kim S -J \emph{et.al.} 2004 \emph{Physica C} {\bf 412-414} 1401
\bibitem{irie05}Irie A and Oya G 2005 \emph{IEEE. Trans. Appl. Supercond.} {\bf 15} 813
\bibitem{okano06}Okano S, Irie A and Oya G 2006 \emph{Journal of the Korean Physical Society} {\bf 48} 1080
\bibitem{komissinski02} Komissinski P V, Il'ichev E, Ovsyannikov G A, Kovtonyuk S A, Grajcar M, Hlubina R, Ivanov Z, Tanaka Y, Yoshida N, Kashiwaya S 2002 \emph{Europhys. Lett.} {\bf 57} 585
\bibitem{matsumoto99} Matsumoto H, Sakamoto S, Wajima F, Koyama T and Machida M 1999 \emph{Phys. Rev. B}  {\bf 60} 3666
\bibitem{krasnov02} Krasnov V M 2002 \emph{Physica C} {\bf 368} 246
\bibitem{gross13}Gross B, Yuan J, An D Y, Li M Y, Kinev N, Zhou X J, Ji M, Huang Y, Hatano T, Mints R G, Koshelets V P, Wu P H, Wang H B, Koelle D and Kleiner R 2013 \emph{Phys. Rev. B} {\bf 88} 014524
\bibitem{koyama96} Koyama T and Tachiki M 1996 \emph{Phys. Rev. B} {\bf 54} 16183
\bibitem{sms-physC06} Shukrinov Yu M, Mahfouzi F and Seidel P 2006 \emph{Physica C} {\bf 449} 62
\bibitem{ryndyk98} Ryndyk D A 1998 \emph{Phys. Rev. Lett.} {\bf80} 3376
\bibitem{iso-apl08} Irie A, Shukrinov Yu M and Oya G 2008 \emph{Appl. Phys. Lett.} {\bf 93} 152510
\bibitem{sr-jetp12} Shukrinov Yu M and Rahmonov I R 2012 \emph{JETP} {\bf 115} 289; \emph{ZhETF} {\bf 142} 323

\bibitem{seidel01} Seidel P, Grib A N, Shukrinov Yu M, Scherbel J, Huebner U, Schmidl F 2001 \emph{Physica C} {\bf 362} 102


\bibitem{bul96} Bulaevskii L N, Dominguez D, Maley M, Bishop A and Ivlev B 1996 \emph{Phys. Rev. B} {\bf 53} 14601
\bibitem{machida99} Machida M, Koyama T and Tachiki M 1999 \emph{Phys. Rev. Lett.} {\bf 83} 4816
\bibitem{ryndyk} Ryndyk D A, Keller J and Helm C 2002 \emph{J. Phys.: Condens. Matter} {\bf 14} 815
\bibitem{helm} Helm Ch, Preis Ch, Forsthofer F, Keller J, Schlenga K, Kleiner R and M$\ddot{u}$ller P 1997 \emph{Phys. Rev. Lett.} {\bf 79} 737;
 Preis Ch, Helm Ch, Schmalzl K, Keller J, Kleiner R and M$\ddot{u}$ller P 2001 \emph{Physica C} {\bf 362} 51
\bibitem{shu-sar02}Shukrinov Yu M, Nasrulaev Kh, Sargolzaei M, Oya G and Irie A 2002 \emph{Supercond. Sci. Technol.} {\bf 15} 178
\bibitem{smp-prb07}Shukrinov Yu M, Mahfouzi F, Pedersen N F 2007 \emph{Phys. Rev. B} {\bf 75} 104508

\bibitem{nakamura09} Nakamura H, Machida M, Koyama T and Hamada N 2009 \emph{J. Phys. Soc. Japan} {\bf 78} 123712
\bibitem{kashiwaya10} Kashiwaya H, Shirai K, Matsumoto T, Shibata H, Kambara H, Ishikado M, Eisaki H, Iyo Y, Shamoto S, Kurosawa I, Kashiwaya S 2010 \emph{Appl. Phys. Lett.} {\bf 96} 202504
\bibitem{ayukawa14} Ayukawa S-Y, Kitano H, Noji T, Koike Y 2014 \emph{JPS Conf. Proc.} 012123
\end{thebibliography}
\end{document}